\documentclass[twocolumn,showpacs,preprintnumbers,amsmath,amssymb]{revtex4}
\usepackage{graphicx}% Include figure files
\usepackage{dcolumn}% Align table columns on decimal point
\usepackage{amsfonts}
\usepackage{mathrsfs}
\usepackage{lscape}
\usepackage{color}

\usepackage{bm}% bold math

\begin{document}

\title{Fourier transform spectroscopy and coupled-channel deperturbation
treatment of the $A^1\Sigma^+ \sim b^3\Pi$ complex of KCs molecule}

\author{A. Kruzins, I. Klincare, O. Nikolayeva, M. Tamanis, R. Ferber}
\affiliation{Laser center, University of Latvia, Rainis Boulevard
19, LV-1586 Riga, Latvia}
\author{E. A. Pazyuk,  A. V. Stolyarov}
\affiliation{Department of Chemistry, Moscow State University, GSP-2
Leninskie gory 1/3, Moscow 119992, Russia }

\date{\today}% It is always \today, today, but any date may be explicitly specified

\begin{abstract}
The laser induced fluorescence (LIF) spectra $A^1\Sigma^+\sim
b^3\Pi(E^J)\to X^1\Sigma^+$ of KCs dimer were recorded in near
infrared region by Fourier Transform Spectrometer with a resolution
of 0.03 cm$^{-1}$. Overall more than 200 collisionally enhanced LIF
spectra were rotationally assigned to $^{39}$K$^{133}$Cs and
$^{41}$K$^{133}$Cs isotopomers yielding with the uncertainty of
0.003-0.01 cm$^{-1}$ more than 3400 rovibronic term values of the
strongly mixed singlet $A^1\Sigma^+$ and triplet $b^3\Pi$ states.
Experimental data massive starts from the lowest vibrational level
$v_A=0$ of the singlet and nonuniformly cover the energy range
$E^J\in [10040,13250]$ cm$^{-1}$ with rotational quantum numbers
$J\in [7,225]$. Besides of the dominating regular $A^1\Sigma^+\sim
b^3\Pi_{\Omega = 0}$ interactions the weak and local heterogenous
$A^1\Sigma^+\sim b^3\Pi_{\Omega = 1}$ perturbations have been
discovered and analyzed. Coupled-channel deperturbation analysis of
the experimental $^{39}$K$^{133}$Cs $e$-parity termvalues of the
$A^1\Sigma^+\sim b^3\Pi_{\Omega = 0,1,2}$ complex was accomplished
in the framework of the phenomenological $4\times 4$ Hamiltonian
accounting implicitly for regular interactions with the remote
$^1\Pi$ and $^3\Sigma^+$ states manifold. The resulting diabatic
potential energy curves of the interacting states and relevant
spin-orbit coupling matrix elements defined analytically by Expanded
Morse Oscillators model reproduce 95\% of experimental data field of
the $^{39}$K$^{133}$Cs isotopomer with a standard deviation of 0.004
cm$^{-1}$ which is consistent with the uncertainty of the
experiment. Reliability of the derived parameters was additionally
confirmed by a good agreement between the predicted and experimental
termvalues of $^{41}$K$^{133}$Cs isotopomer. Calculated relative
intensity distributions in the $A\sim b \to X$ LIF progressions are
also consistent with their experimental counterparts. Finally, the
deperturbation model was applied for a simulation of pump-dump
optical cycle $a^3\Sigma^+ \to A^1\Sigma^+\sim b^3\Pi \to
X^1\Sigma^+$ proposed for transformation of ultracold colliding
K$+$Cs pairs to their ground molecular state $v_X=0;J_X=0$.
\end{abstract}
\maketitle

\section{Introduction}
In ultracold (below 1 mK) gases of polar alkali diatomic molecules
the electric dipole-dipole interaction possesses long-range and
anisotropic character and cause particular attraction since they are
promising for a variety of novel applications \cite{hutson}. Among
them there are proposals of quantum phase transitions \cite{qpt},
quantum information devices \cite{qid}, coherent control of chemical
reactions \cite{chem react}. Further success in producing ensembles
of ultracold quantum gases of diatomic molecules relies on the
accurate knowledge of the molecular rovibronic structure and
transition probabilities gained from the high resolution
spectroscopy and state of art \emph{ab initio} calculations.

\begin{figure}
%\begin{center}
\includegraphics[scale=1.0]{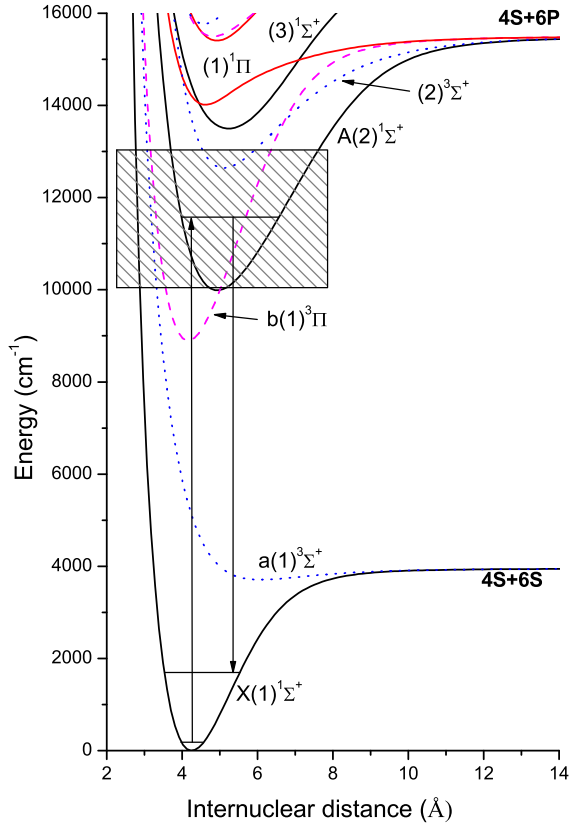}
\caption{(Color online) Scheme of electronic terms based on \emph{ab
initio} results of Ref.\ \cite{korek-00}}\label{PECab}
%\end{center}
\end{figure}

As far as polar alkali diatomic molecules are considered, experiment
based information on the potential energy curves of the lowest
excited $A^1\Sigma^+$ and $b^3\Pi$ states mixed by singlet-triplet
spin-orbit (SO) interaction is of great importance due to the
possibility of their usage as intermediate states for transferring
the vibrationally excited molecules obtained from cold colliding
atoms via photoassociation or Feshbach resonances into the absolute
rovibronic ground states $X^1\Sigma^+$ with $v_X=0,J_X=0$
\cite{stw-04,ghosal-09}. In the sequence of heteronuclear alkali
diatomics KCs is among promising species for producing ultracold
polar quantum gases. Due to proximity of K$(4p)$ and Rb$(5p$)
energies, the energy of low-lying electronic states of KCs is
similar to the ones in the RbCs molecule, however, KCs possess
larger permanent electric dipole moment. The RbCs molecule has been
already studied in ultracold conditions
\cite{sage-2005,sage-2004,pilch-2009}. In particular, the recent
quantum dynamics simulation of the mixed $A^1\Sigma^+$ and
$b^3\Pi_{\Omega}$ states of RbCs in a time-dependent wave-packet
approach \cite{ghosal-09} demonstrated that the pump - dump
picosecond pulsed laser scheme could be efficient to form ultracold
heteronuclear diatomic molecules in highly bonded ground
$X^1\Sigma^+$ state. The highly accurate empirical potentials for
the ground singlet $A^1\Sigma^+$ and triplet $a^3\Sigma^+$ states
converging to the lowest K(4S)$+$Cs(6S) asymptote (see Fig.\
\ref{PECab}) as well as predicted scattering lengths and Feshbach
resonances which are required to simulate cold collision processes
are presented in Ref's \cite{KCs-ground,KCs-triplet}. However, till
now there is no empirical information about the excited states of
KCs while \emph{ab initio} potential energy curves (PECs) have been
calculated for a wide range of internuclear distance in the
framework of pure Hund's (\textbf{a})
\cite{korek-00,Aymar2008,KS-09} and (\textbf{c}) \cite{korek-06}
coupling cases. The relevant permanent and transition dipole moment
functions are also available from Ref's \cite{Aymar2008,KS-09}.

Particular complicated objects for the theoretical interpretation
are alkali diatomics containing a heavy Rb or Cs atoms since the
relevant singlet-triplet SO coupling matrix element
$\xi^{so}_{Ab_0}$ between $A^1\Sigma^+$ and $b^3\Pi_{\Omega=0}$
states is comparable with vibrational spacing of the interacting
states. Therefore $A^1\Sigma^+$ and $b^3\Pi_{\Omega=0}$ states are
fully mixed (later denoted as the $A \sim b$ complex) in both
adiabatic and diabatic basis set representation \cite{field}. First
studies of such kind of system have been reported for the
Rb-containing alkali diatomics, in which apparent disorder in
vibrational spacing was observed for Rb$_2$ \cite{Amiot-rb2}, see
\cite{Salami-09} for more extensive studies, NaRb \cite{narb-02} and
RbCs \cite{Bergeman-2003}. The crucial issue however remained
unsolved, namely, while the experimental term values were obtained
with high precision of about 0.003-0.01 cm$^{-1}$, the resulting
deperturbed parameters reproduced experimental data with much poorer
accuracy of 0.05-0.25 cm$^{-1}$ \cite{narb-02, Bergeman-2003,
Salami-09}. Furthermore, even the vibrational numbering of the
``dark" triplet $b^3\Pi$ state was still remained questionable.

In recent studies on the $A \sim b$ complex of NaRb
\cite{Docenko-2007} and NaCs \cite{Zaharova-2009} more comprehensive
deperturbation method has been developed in the framework of the
inverted channel-coupling approach by means of the $4 \times4 $
Hamiltonian for the $A^1\Sigma^+$ and $b^3\Pi_{\Omega = 0,1,2}$
substates constructed on Hund's coupling case (a) basis functions.
Phenomenological inclusion of the splitting of the $b^3\Pi_{\Omega =
0,1,2}$ substates and of indirect $A^1\Sigma^+$ and $b^3\Pi_{\Omega
= 1}$ states coupling allowed accounting for the regular
perturbation. The elaborated model reproduced the experimental term
values obtained by the high resolution Fourier transform
spectroscopy (FTS) of the laser induced fluorescence (LIF) from/to
the $A \sim b$ complex with a standard deviation of 0.006-0.012
cm$^{-1}$, which was consistent with the uncertainty of the FTS
experiment. The paper \cite{Zaharova-2009} presents the most
challenging though attractive, both for model's testing and
applications, case of $A \sim b$ complex study of a Cs-atom
containing  alkali diatomics where SO interaction is of largest
value mainly determined by $\xi_{\rm so} \sim 185$ cm$^{-1}$ for
$6^2P$(Cs). This successful approach opens the path to study of the
$A \sim b$ complex of more heavier Cs-containing heteronuclear
alkali diatomic such as RbCs and KCs. While the RbCs molecule is
currently under investigation \cite{Docenko_Ab_RbCs} the present
paper deals with the KCs molecule.

It should be noted that the present experimental and deperturbation
studies of the KCs $A \sim b$ complex are greatly facilitated by the
couple of facts. First, the KCs molecule was recently involved in
detailed FTS LIF investigation \cite{KCs-ground} which resulted in
high accuracy empirical point-wise PEC for the ground $X^1\Sigma^+$
state covering a range $v_X = 0$ to 97 and $J_X = 12$ to 209.
Moreover, a majority of the data for PEC construction in
\cite{KCs-ground} was obtained directly from the $(A \sim b)
\rightarrow X$ LIF spectra containing the information on rovibronic
energies of the $A \sim b$ complex. Second, the accurate \emph{ab
initio} spin-orbit coupling matrix elements between the lowest
electronic states of KCs have been obtained recently in Ref.\
\cite{KS-09} as explicit functions of internuclear distance.

The goal of the present work is to perform high-resolution
spectroscopy studies of the $A \sim b$ complex of the KCs molecule
and to apply the adequate data processing in order to get empirical
molecular structure parameters, namely the deperturbed $A^1\Sigma^+$
and $b^3\Pi$ diabatic potential energy curves along with the related
SO coupling matrix elements functions on internuclear distance $R$.

In summary, Sections II and III present the experimental techniques
and spectra analysis, Section IV presents the deperturbation model
and computational details, Section V includes obtained fitted
parameters, as well as interpretation of experimental data. The
paper ends with short conclusion (Section VI).

\section{Experiment}

In the experiment the rovibronic levels of the KCs $A \sim b$
complex were directly exited by diode lasers and subsequent $(A \sim
b) \rightarrow X$ LIF spectra were recorded by Fourier transform
spectrometer. The experimental setup was the same as used for KCs
ground state studies \cite{KCs-ground} and, hence, we will mention
here only some key points of the experiment. KCs molecules were
produced in a linear stainless steel heat pipe. The heat pipe was
filled with 10 g of K (natural isotope mixture) and 7 g of Cs (both
metals in ampules from Alfa Aesar). The typical operating pressure
of Ar buffer gas in the heat pipe was 3-5 mbar. On the one hand Ar
gas prevented condensation of the metal vapor on the heat pipe
windows, and, on the other  hand, collisions of optically excited
KCs molecules in particular rovibronic levels $v', J'$  with Ar
atoms ensured population of great number of neighboring rotational
levels and observation  of the respective transitions to the ground
state. During the experiments the heat pipe was  kept at $270 -
290^{\rm o}$C temperature by a Carbolite furnace.

In the experiment the laser beam was sent into the heat pipe through
a pierced mirror. Backward LIF was collected by the same mirror and
focused, by two lenses, on the input aperture of the spectrometer
(Bruker IFS 125HR). For LIF detection we used an InGaAs diode
operated at room temperature. The spectral sensitivity of the
detector gradually diminishes from about 6500 cm$^{-1}$ toward
higher frequencies reaching 60\% of maximal sensitivity at $\sim
10000$ cm$^{-1}$. Then sensitivity drops much faster, reaching
10-15\% of maximal value at 11000 cm$^{-1}$. The resolution of the
spectrometer was typically set to 0.03 cm$^{-1}$.

The $(A \sim b) \leftarrow X$ transitions were excited by tunable
diode lasers. Four single mode laser diodes with central wavelenghts
850 nm (LD850/100 from Toptica Photonics), 980 nm (L980P200I from
Thorlabs), 1020 nm (LD-1020-0400 from Toptica Photonics) and 1060 nm
(L1060P100J from Thorlabs) were mounted in homemade external cavity
resonators (Littrow configuration) with a grating serving as a
feedback source. The respective frequency tuning ranges were 11560 -
11930 cm$^{-1}$, 10209 - 10515 cm$^{-1}$, 9700-9860 cm$^{-1}$, and
9360-9510 cm$^{-1}$. Fine tuning of the grating was achieved by a
piezoelectric actuator. Temperature and current stabilization was
ensured by Thorlabs controllers. The power of the lasers at the
entrance of the heat pipe varied from 15 to about 50 mW depending on
particular laser diode and laser current. The general approach to
measurements was the following. The laser frequency was tuned until
the LIF signal monitored at reduced resolution in the ``Preview
Mode" of the spectrometer gets its maximal value and then fixed
during recording of the spectrum. In order to ensure sufficient
signal-to-noise ratio for the lines of medium strength, the number
of scans for each recorded spectrum varied from 20 to 40; averaging
over a number of repeated measurements was applied in some cases. In
some cases the strong scattered laser light from the input window of
the heat pipe was eliminated by long-pass edge filters.

\begin{figure}[t!]
%\begin{left}
\includegraphics[scale=0.7]{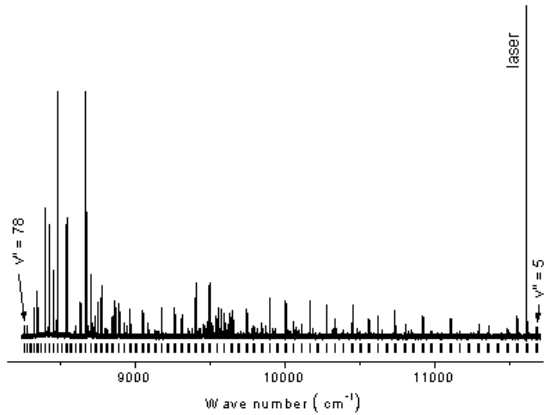}
\caption{$^{39}$K$^{133}$Cs $(A \sim b) \rightarrow X$ LIF spectrum
recorded by InGaAs detector at excitation with 850nm laser diode
(laser frequency 11608.623 cm$^{-1}$). The strongest doublet
$P_{63}, R_{61}$ progression $5 \le v_X \le 78$ originates from the
level $J'= 62$ with energy $E'= 12164.984$ cm$^{-1}$ excited by the
transition $(v^*, 62) \leftarrow (6, 63)$. Another weaker KCs LIF
progression excited in $(v^*, 133) \leftarrow (6, 134)$ transition
is not marked. A mesh of weaker lines around 9500 cm$^{-1}$ belongs
to K$_2$ LIF.} \label{spec1}
%\end{left}
\end{figure}

\section{Analysis of the Spectra}

Recorded KCs LIF spectra, see Figs.\ 2-5, typically consisted of one
or two strong doublet progressions and several weaker ones.

The assignment of the LIF progressions in the $(A \sim b)
\rightarrow X$
 spectra was straightforward from measured vibrational and rotational
 spacings thanks to accurate ground state PEC \cite{KCs-ground}.
 The term values of the upper rovibronic levels of the
 $A \sim b$ complex giving rise to $(A \sim b) \rightarrow X$ LIF were obtained by adding the
 corresponding ground state level energy to a transition wave number.
The uncertainty of the line positions is estimated to be 0.1 of the
resolution, or 0.003 cm$^{-1}$. For lines with signal-to-noise ratio
(SNR) less than 3 the uncertainty increases and was assumed to be
about 0.01 cm$^{-1}$ for lines with SNR of about 2. The term values
of the upper state levels were determined as average value from at
 least four transitions. Hence, the main uncertainty of the term values could be
 caused by the shift (if any) of the laser frequency from the center of the Doppler
 broadened $(A \sim b) \leftarrow X$ absorption line. When LIF is
 registered, as in our case, along the direction of the laser beam,
 this shift is transferred to the respective shift of the LIF line
 positions. The Doppler broadening for KCs at the exploited spectral
 region is about 0.012 cm$^{-1}$ and we estimate maximal
 experimental uncertainty of the $A \sim b$ complex term values as
 0.01 cm$^{-1}$.

 Typically, around the strong LIF lines the satellite lines could be
recorded due to collision - induced distribution of the population
of the directly excited rovibronic level over neighboring rotational
levels and even vibrational levels. The assignment of these lines
was based on the high accuracy of the ground state PEC and it was
always checked that the spacing of satellite doublet lines $(v_R -
v_P)$ coincides with calculated energy differences between the
respective ground state rotational levels within accuracy of some
miliwavenumbers.

Figure 2 represents an example of the recorded LIF spectrum when the
$(A \sim b) \leftarrow X$   transition is excited with the 850 nm
laser diode tuned to 11608.623 cm$^{-1}$. The recorded spectral line
intensities for frequencies above 10000 cm$^{-1}$ are substentially
diminished because of dropping of the spectral sensitivity of the
InGaAs detector. The LIF intensity distribution in such a long
progression as depicted in Fig.\ 2, with strong last maximum at high
$v_X$, is characteristic for highly excited vibrational levels $v_A$
of the $A \sim b$ complex. Note that, in general, the LIF spectra
recorded with 850 nm diode excitation usually contained a very
strong $A^1\Sigma_u \rightarrow X^1\Sigma_g$ band of K$_2$, and only
by careful selection of the laser frequencies it was possible to
eliminate the K$_2$ fluorescence, at the same time having strong
enough KCs $(A \sim b) \rightarrow X$ LIF progressions.

\begin{figure}
%\begin{center}
\includegraphics[scale=1.0]{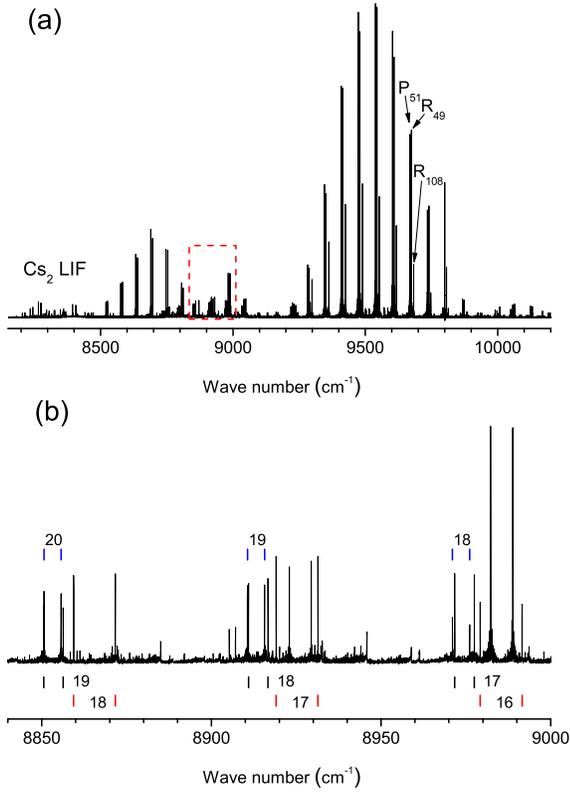}
\caption{(a) The $^{39}$K$^{133}$Cs $(A \sim b) \rightarrow X$ LIF
spectrum recorded at laser frequency 9800.135 cm$^{-1}$. The
following $(A \sim b) \leftarrow X$; $(v^*, J') \leftarrow (v_X,
J_X)$ excitation transitions have been assigned: $(v^*,50)
\leftarrow (4,51)$; $(v^*,109) \leftarrow (3,110)$; $(v^*, 57)
\leftarrow ( 7,56)$; $(v^*,123) \leftarrow (16,122)$; $(v^*,44)
\leftarrow (5,43)$. Weak Cs$_2$ progressions have been assigned
below 8500 cm$^{-1}$. An indexes for $P_{J_X},R_{J_X}$ lines denote
the respective ground state rotational levels.
 (b) The zoomed fragment of the spectrum;  doublets originating
 from the lowest excited vibrational level posessing dominant singlet character
 with $J'= 50$ and $J'=109$
  are marked with black and red
 vertical bars below the spectrum. The bars above the spectrum
 mark transitions from the level $(v^*, J'=44)$ with  dominant triplet character.
 Numbers at bars denote $v_X$.} \label{spectr2}
%\end{center}
\end{figure}

The usage of 1020 nm and 1050 nm laser diodes allowed us to reach
low lying $A$ - state rovibronic levels of KCs. Rather strong LIF
signal made it possible in a number of cases to detect spectra
without filtering of scattered laser light. Thus, full LIF
progressions could be recorded, which was important for establishing
the relative intensity distribution in the progressions, which is
helpful for vibrational assignment of the upper state. The example
of such a spectrum is given in Fig.\ 3. This spectrum contains five
assigned KCs $(A \sim b) \rightarrow X$ progressions, two of them
with $J' = 50$ and 109 are originating from the lowest $v_A$ level
with predominant singlet character. These two progressions show
unusual intensity distribution. Along with one strong bell - shape
maximum in intensity distribution around 9500 cm$^{-1}$ at low
$v_X$, as is typical for progressions from upper state $v' = 0$
vibrational level, we have found for these progressions transitions
to higher $v_X$ resulting in additional intensity maximum, however
substantially weaker than the main one. The small fragment of the
spectrum with these additional transitions is given in Fig.\ 3b.
Besides, the spectrum in Fig.\ 3 contains a progression originating
from the level with energy $E' = 10227.038$ cm$^{-1}$ and $J'= 44$
which is of predominantly triplet character as will be shown below,
see Section V. A short fragment of this progression can be seen in
Fig.\ 3b where the respective $P, R$-doublets are marked above the
spectrum. In general, the laser most often exited the levels with a
dominant singlet character, and only in some cases a direct
excitation of rovibronic $(A \sim b)$-state levels with a dominant
triplet character took place.

\begin{figure}
%\begin{center}
\includegraphics[scale=1.0]{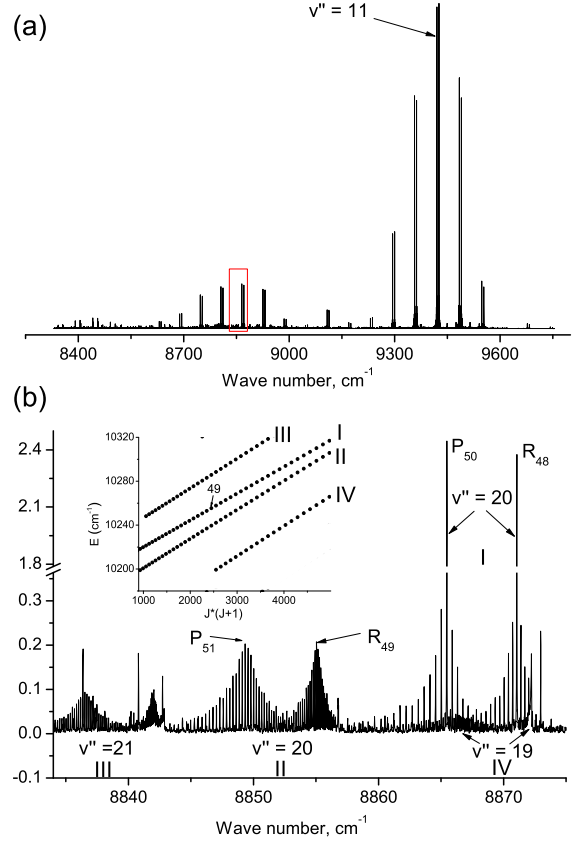}
\caption{(a) The $^{39}$K$^{133}$Cs $(A \sim b) \rightarrow X$ LIF
spectrum recorded at excitation frequency 9743.083 cm${-1}$. The
scattered laser light and LIF at frequencies higher than about 9500
cm${-1}$ are cut by a long-pass edge filter. The following
excitation transitions for $^{39}$K$^{133}$Cs have been assigned:
$(v^*,49)\leftarrow(6,50); (v^*,81)\leftarrow (17,82); (v^*, 25)
\leftarrow (13,26); (v^*,135)\leftarrow(12,134); (v^*,150)\leftarrow
(11,149)$. (b) Zoomed part of the spectrum around $v_X= 20$ for the
strongest progression with $J'= 49$. Three groups (II-IV) of
satellite $P,R$ branches shifted from the main rotational relaxation
pattern (group I) are clearly seen. The inset presents term values
of the respective rovibronic levels as dependent of $J(J+1)$.
Directly excited level $J' = 49$ ir marked.} \label{CIF}
%\end{center}
\end{figure}

\begin{figure}
%\begin{center}
\includegraphics[scale=1.0]{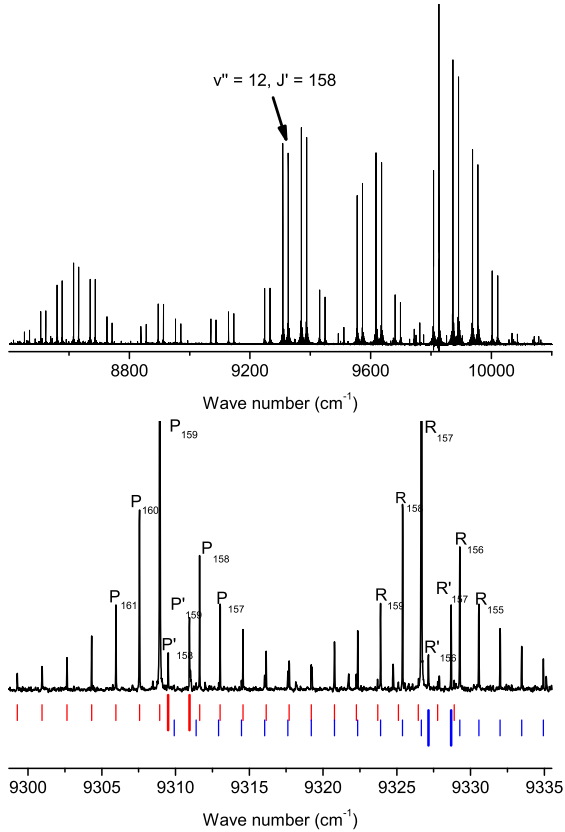}
\caption{(Color online) (a) The LIF spectrum recorded at excitation
frequency 9826.348 cm$^{-1}$. The strongest progression originates
from the upper level excited in transition $(v^*, 158) \leftarrow
(4,157)$. (b) zoomed part of the spectrum around $v_X=12$; Red and
blue bars denote assigned P and R satellite lines respectively. Long
bars mark transitions from upper state levels with dominant triplet
$b^3\Pi_1$ character. Numbers in the spectrum denote corresponding
ground state rotational quantum numbers for observed transitions.}
\label{loc_pert}
%\end{center}
\end{figure}

Fig.\ 4 and Fig.\ 5 demonstrate how fruitful is the collision
induced rotational relaxation for systematic study of the $A \sim b$
complex level structure. The spectrum in Fig.\ 4a contains one very
strong progression originating from the level $E'= 10255.413$
cm$^{-1}$, $J^{\prime}= 49$ as well as five weaker ones. A large
number of satellite lines was assigned around the strong $P_{50},
R_{48}$ doublet lines, especially at transitions to $v_X=11$. The
most interesting feature consists in appearance of the additional
three wide groups of $P, R$ satellite lines, as seen in Fig.\ 4b.
These groups are shifted from the main $P_{50}, R_{48}$ doublet and
represent a situation when an initial population of the optically
excited level with $J^{\prime}= 49$ is collisionally distributed
within two state ($A^1\Sigma^+$ and $b^3\Pi_0$) rovibronic level
manifolds of the $A \sim b$ complex. As a result, more than 130 term
values were obtained from this spectrum, see the inset in Fig.\ 4b.
Note that these satellite groups are the strongest in the range 8700
cm$^{-1}$ -- 9000 cm$^{-1}$ and substantially weaker at higher
frequency (lower $v_X$) range.

In several spectra clear evidences of local perturbations were
observed as illustrated in Fig.\ 5. The spectrum given in Fig.\ 5a
is recorded at excitation frequency 9826.348 cm$^{-1}$; it consists
of one strong KCs LIF progression from the level ($E' = 10861.046$
cm$^{-1}$, $J^{\prime} = 158$) and several weak KCs and K$_2$ LIF
progressions, hardly seen in this scale. Zoomed part of the strong
progression in the range of $v_X=12$ is given in Fig.\ 5b. Along
with an ordinary satellite line pattern around the central $P, R$
lines with gradually changing spacing between the lines, the local
perturbation can be clearly recognized from the two gaps
(``windows") in the both $P, R$ branches, see $P_{159}/P_{158}$ and
$R_{157}/R_{156}$. Moreover, we have assigned ``extra" lines with
the same rotational quantum numbers, namely $P'_{158}, P'_{159},
R'_{156}$ and $R'_{157}$, see Fig.\ 5b. As it will be shown later
these additional lines take their origin from the levels with
dominant $b^3\Pi_1$ state character and they can be seen here due to
effective collisional transfer of population from the directly
excited predominantly singlet $A$ state level with $J^{\prime}=
158$, since it is very close to the perturbation center.

Overall such local perturbations accompanied with extra lines were
observed in three spectra. In several cases the local perturbations
were observed as irregularities in line sequences without ``extra"
lines.

\begin{figure}
%\begin{center}
\includegraphics[scale=0.7]{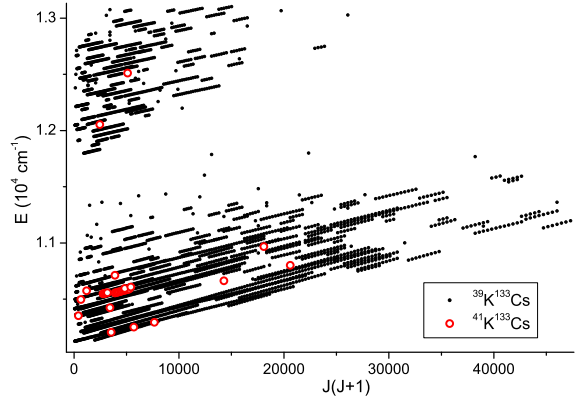}
\caption{(Color online) Experimental term values of the
$(A^1\Sigma^+ \sim b^3\Pi)$ complex plotted versus
$J^{\prime}(J^{\prime}+1)$.} \label{gero_plot}
%\end{center}
\end{figure}

Summarizing, during this study more than 3400 term values of the $A
\sim b$ complex rovibronic levels were obtained. These term values
are given in Fig.\ \ref{gero_plot} as dependent on
$J^{\prime}(J^{\prime}+1)$. Empty circles seen in Fig.\
\ref{gero_plot} show distribution of 31 term value determined for
the $^{41}$K$^{133}$Cs levels. The data for $^{41}$K$^{133}$Cs are
given also in Table~\ref{isotope}.

%******************************** Table IV ********************************
\begin{table}[t!]
\caption{The experimental rovibronic termvalues $E^{expt}$ (in
cm$^{-1}$) of the of $^{41}$K$^{133}$Cs isotopomer assigned to the
$A^1\Sigma^+ \sim b^3\Pi$ complex. $\Delta=E^{expt}-E^{CC}$ is the
difference between experimental and predicted energies. $P_i$ (in
$\%$) are fractional partition of the levels, see Section
V.}\label{isotope}
\begin{center}
\begin{tabular}{cccccc}
\hline \hline
$J^{\prime}$ & $E^{expt}$ & $\Delta$ & $P_A$ & $P_{b0}$ & $P_{b1}$\\
\hline
 59 & 10205.767 &  0.003 &  44.9 &  55.1 & 0.0\\
 75 & 10250.792 & -0.001 &  85.2 &  12.7 & 2.1\\
 58 & 10424.388 &  0.001 &  65.8 &  34.2 & 0.0\\
 52 & 10547.339 & -0.008 &  66.3 &  33.5 & 0.2\\
 53 & 10549.850 & -0.008 &  66.8 &  32.9 & 0.3\\
 54 & 10552.405 & -0.007 &  67.2 &  32.3 & 0.5\\
 55 & 10555.010 &  0.000 &  67.5 &  31.6 & 0.9\\
 56 & 10557.639 & -0.014 &  67.5 &  30.7 & 1.8\\
 62 & 10555.967 & -0.003 &  51.1 &  48.8 & 0.1\\
 63 & 10559.120 &  0.001 &  50.4 &  49.5 & 0.1\\
 64 & 10562.317 & -0.007 &  49.7 &  50.2 & 0.1\\
 65 & 10565.579 & -0.007 &  49.0 &  50.9 & 0.1\\
 66 & 10568.900 & -0.006 &  48.3 &  51.6 & 0.1\\
 67 & 10572.278 & -0.005 &  47.6 &  52.4 & 0.1\\
 68 & 10575.711 & -0.006 &  46.8 &  53.1 & 0.1\\
 69 & 10579.199 & -0.010 &  46.1 &  53.8 & 0.1\\
 70 & 10582.754 & -0.007 &  45.3 &  54.6 & 0.1\\
 71 & 10586.355 & -0.014 &  44.6 &  55.3 & 0.1\\
 72 & 10590.031 & -0.006 &  43.9 &  56.1 & 0.1\\
 69 & 10595.554 & -0.006 &  75.6 &  24.3 & 0.1\\
 62 & 10713.205 & -0.006 &  61.5 &  38.4 & 0.1\\
 25 & 10496.360 & -0.011 &  55.7 &  44.3 & 0.0\\
 73 & 10608.598 &  0.005 &  77.4 &  22.6 & 0.1\\
 34 & 10575.252 &  0.002 &  83.9 &  16.0 & 0.0\\
 87 & 10294.767 &  0.004 &  86.3 &  13.6 & 0.1\\
 20 & 10352.114 &  0.004 &  61.3 &  38.6 & 0.0\\
119 & 10662.753 & -0.001 &  77.3 &  22.5 & 0.1\\
143 & 10801.142 & -0.001 &  70.9 &  27.4 & 1.8\\
134 & 10968.075 &  0.005 &  80.2 &  19.5 & 0.3\\
 49 & 12053.926 & -0.001 &  82.7 &  17.3 & 0.0\\
 71 & 12511.108 &  0.009 &  81.7 &  18.3 & 0.0\\
\hline
\end{tabular}
\end{center}
\end{table}

\section{Deperturbation analysis}\label{model}

\subsection{Modeling Hamiltonian}

The non-adiabatic rovibronic wavefunction
$\Psi_j=\sum_{i}\phi_i\varphi_i$ ($i\in [A^1\Sigma^+,
b^3\Pi_{\Omega=0,1,2}]$) corresponding to the $j$-th rovibronic
level of the $A^1\Sigma^+\sim b^3\Pi$ complex of KCs molecule with
the fixed rotational quantum number $J$ and $e$-symmetry was
approximated by the linear combination of the symmetrized
electronic-rotational wavefunctions $\varphi_i$ belonging to a pure
Hund's coupling case (\textbf{a}) \cite{field}. The expansion
coefficients $\phi_i(r)$ are the $r$-dependent fractional components
of the non-adiabatic vibrational wavefunction $\mathbf{\Phi}_j\equiv
||\phi_A, \phi_{b0}, \phi_{b1}, \phi_{b2}||$ which are determined by
the bound solution of the close-coupled (CC) radial equations \cite{
Meshkov05, Docenko-2007, Zaharova-2009}:
\begin{eqnarray}\label{CC}
\left(- {\bf I}\frac{\hbar^2 d^2}{2\mu dr^2} + {\bf V}(r;\mu,J) -
{\bf I}E^{CC}_j\right)\mathbf{\Phi}_j(r) = 0
\end{eqnarray}
with the conventional boundary $\phi_i(0)=\phi_i(\infty)=0$ and
normalization $\sum_{i}P_i=1$ conditions, where $
P_i=\langle\phi_i|\phi_i\rangle$ is the fractional partition of the
$j$-th level. Here ${\bf I}$ is the identity matrix, $E^{CC}_j$ is
the total non-adiabatic energy of rovibronic level of the complex
while ${\bf V}$ is the symmetric $4\times 4$ matrix of potential
energy consists of the diagonal
\begin{eqnarray}\label{modeldAb4}
V_{^1\Sigma^+} &=& U_A+B[X+2]\\
V_{^3\Pi_0} &=& U_{b0}+B[X+2]\nonumber\\
V_{^3\Pi_1} &=& U_{b1}+B[X+2]\nonumber\\
V_{^3\Pi_2} &=& U_{b2}+B[X-2]\nonumber
\end{eqnarray}
and non-vanishing off-diagonal
\begin{eqnarray}\label{modelodAb4}
V_{^1\Sigma^+-^3\Pi_0}&=&-\sqrt{2}\xi^{so}_{Ab0}\\
V_{^3\Pi_0-^3\Pi_1}   &=&-B\sqrt{2X}\nonumber\\
V_{^3\Pi_1-^3\Pi_2}   &=&-B\sqrt{2(X-2)}\nonumber\\
V_{^1\Sigma^+-^3\Pi_1}&=&-B\zeta_{Ab1}\sqrt{2X}\nonumber
\end{eqnarray}
matrix elements explicitly depending on reduced mass $\mu$ and
rotational quantum number $J$ as a parameter:
\begin{eqnarray}\label{BJ}
B\equiv\frac{\hbar^2}{2\mu r^2}; \quad X\equiv J(J+1)\nonumber
\end{eqnarray}
Hereafter all electronic parameters of the model are assumed to be
mass-invariant.

Here $U_A(r)$, $U_{b\Omega}(r)$ are the diabatic PECs of the singlet
$A^1\Sigma^+$ and triplet $b^3\Pi_{\Omega = 0,1,2}$ sub-states while
$\xi^{so}_{Ab_0}(r)$ is the relevant spin-orbit coupling matrix
element. Besides of strong off-diagonal homogenous $A^1\Sigma^+\sim
b^3\Pi_{\Omega = 0}$ interactions the deperturbation model also
explicitly consider heterogenous spin-rotational interaction between
all $\Omega=0,1,2$ components of the triplet $b^3\Pi$ state
\cite{field}. Moreover, the non-equidistant SO splitting functions
$A^{so}_{-}\neq A^{so}_{+}$ between the $b^3\Pi_{\Omega=0,1,2}$
sub-states
\begin{eqnarray}\label{ASO}
U_{b1}=U_{b0}+A^{so}_{-};\quad U_{b2}=U_{b1}+A^{so}_{+}
\end{eqnarray}
and the indirect coupling $A^1\Sigma^+\sim b^3\Pi_{\Omega=1}$
parameter $\zeta_{Ab1}$ were introduced in the Hamiltonian
(\ref{modeldAb4}); (\ref{modelodAb4}) in order to account implicitly
for regular perturbations caused by the remote $^1\Pi$ and
$^3\Sigma^+$ states manifold (first of all the $B^1\Pi$ and
$c^3\Sigma^+$ states converging to the second dissociation limit,
see Fig.\ \ref{PECab}).

\subsection{Computational details/Fitting procedure}

The empirical PECs of the interacting $A^1\Sigma^+$ and $b^3\Pi_0$
states $U_{A}(r)$, $U_{b0}(r)$ as well as both diagonal
$A^{so}_{\pm}(r)$ and the off-diagonal $\xi^{so}_{Ab0}(r)$
spin-orbit functions were represented analytically by the Expanded
Morse Oscillator (EMO) function \cite{EMO}
\begin{eqnarray}\label{EMO}
T_e+{\mathfrak D}_e\left [1 - e^{-\alpha(r)(r-r_e)}\right ]^2;\\
\alpha = \sum_{i=0}^N a_i\left (
\frac{r^p-r_{ref}^p}{r^p+r_{ref}^p}\right)^i\nonumber
\end{eqnarray}
converging to the appropriate atomic limit. In particular, for all
SO functions $T_e^{so}=\xi_{Cs}^{so}-{\mathfrak D}_e^{so}$ where
$\xi_{Cs}^{so} \equiv [E_{6^2{\rm P}_{3/2}}-E_{6^2{\rm P}_{1/2}}]/3$
is the spin-orbit constant of the Cs atom in the $6^2P$ state
\cite{EW70, WS87}. For diabatic PEC $U_A$ of the singlet state
$T_e^A=T_{dis}-{\mathfrak D}_e^A$ while
$T_e^{b0}=T_{dis}-\xi_{Cs}^{so}-{\mathfrak D}_e^{b0}$ for the
$U_{b0}$ sub-state of the triplet. Here $T_{dis}={\mathfrak
D}_e^X+E_{6^2{\rm P}}-E_{6^2{\rm S}}$ is the energy of center of
gravity of the $Cs(6^2{\rm P})$ doublet (without hfs splitting)
\cite{EW70} referred to the minimum of the ground $X$-state. The
${\mathfrak D}_e^X=4069.3$ cm$^{-1}$ value is taken from Ref.\
\cite{KCs-ground}.

The initial EMO parameters (${\mathfrak D}_e$, $r_e$ and $a_i$) of
the relevant PECs were estimated by averaging of the independent
\emph{ab initio} results borrowed from Ref's \cite{korek-00,
Aymar2008, KS-09} while the required SO functions were constructed
using the recent quasi-relativistic estimates of Ref.\ \cite{KS-09}.
Off-diagonal SO function could be empirically determined only in the
close vicinity of the crossing point of the diabatic PECs of the
interacting $A^1\Sigma^+$ and $b^3\Pi_0$ states. Therefore, the
required SO-EMO functions were constrained by fixing of the
respective $a_i(i>0)$ parameters based on the relevant \emph{ab
initio} points \cite{KS-09}, so only ${\mathfrak D}_e^{so}$,
$r_e^{so}$ and $a_0^{so}$ SO-EMO parameters were rest to be variable
in the fit.

The refined parameters of the PECs and SO functions combined with
the $r$-independent $\xi^{so}_{Ab1}=const$ parameter were determined
iteratively during the weighted nonlinear least-squared fitting
(NLSF) procedure procedure \cite{field}:
\begin{eqnarray}\label{chisexp}
\chi^2_{expt}=\sum^{N_{expt}}_{j=1}\frac{w_j(E^{CC}_j-E^{expt}_j)^2}{N_{expt}-M_p},
\end{eqnarray}
where $w_j=1/\sigma_j^2$ is the weight of each level, $E^{expt}_j$
its experimental termvalue and $\sigma_j^{expt}=0.003-0.01$
cm$^{-1}$ its uncertainty. Here, $N_{expt}$ is the number of
experimental termvalues involved while $M_p$ is the total number of
adjusted parameters of the model. Only experimental rovibronic
termvalues $E^{expt}_j$ of the most abundant $^{39}$K$^{133}$Cs
isotopomer were included in the fitting procedure. The minimum of
the functional (\ref{chisexp}) was searched by the modified
Levenberg-Marquardt algorithm \cite{NR} realized by MINPACK software
\cite{MINPACK}.

The analytical mapping procedure based on replacement of the
conventional radial coordinate (in \AA ) by the reduced radial
variable $y(r; \bar{r}=5.2, \beta=5)= [1+(\bar{r}/r)^{\beta}]^{-1}$
was used to transform the initial CC equations (\ref{CC}) into
completely equivalent form. Then, the modified CC equations given
explicitly in Ref.\ \cite{Meshkov1PRA08} were solved on the interval
$r\in[2.5,15.5]$ \AA $ $    by the finite-difference (FD) boundary
value method \cite{truhlar} with the fixed number of uniformed grid
points $N$. The central 5-points FD approximation (FD5) of the
kinetic energy term was employed in Eq.\ (\ref{CC}). The ordinary
eigenvalue and eigenfunction problem of the resulting symmetric band
matrix was iteratively solved by the implicitly restarted Lanczos
method realized in ARPACK software in the shift-inverted spectral
transformation mode \cite{ARPACK}. The energy error correction for
the FD5 method recently invented in the explicit integral form
\cite{Zaharova-2009} was used to extrapolate to infinite number of
grid points the eigenvalue obtained at the fixed number of
integration points. The optimized procedure allowed us to attain the
absolute accuracy of the calculated energies $E^{CC}_j$ about of
0.001 cm$^{-1}$ using only $N=1500-2000$ grid points.

\section{Results and Discussion}

\subsection{Interatomic potentials and spin-orbit functions}\label{resultHamiltonian}
The resulting EMO parameters of the deperturbed (diabatic) PECs of
the $A^1\Sigma^+$ and $b^3\Pi_0$ states are presented Table
\ref{Abpar} while the respective empirical spin-orbit functions are
given in Table \ref{SOpar}. The parameters reproduce 95\% of
experimental data field of the $^{39}$K$^{133}$Cs isotopomer with a
standard deviation (SD) of 0.004 cm$^{-1}$ in the framework of the
phenomenological model defined by Eq.\ (\ref{modeldAb4}) and Eq.\
(\ref{modelodAb4}). The SD value is well-consistent with the
estimated uncertainty of $0.003-0.01$ cm$^{-1}$ of the present
experiment. The EMO parameters defined by Eq.\ (\ref{EMO}) are given
in the EPAPS in double precision format \cite{EPAPS}. For
convenience the fitted PECs and SO functions are represented in the
EPAPS by the appropriate point-wise form as well. Furthermore, EPAPS
tables contain both experimental and reproduced robvibronic
termvalues of the complex as well as their residuals and fractional
partitions.

%******************************** Table I ********************************
\begin{table}[t!]%[H]
\caption{The resulting EMO parameters of the diabatic potential
energy curves of the deperturbed $A^1\Sigma^+$ and $b^3\Pi_0$
states. $T_{dis}$, ${\mathfrak D}_e$ and $T_e$ in cm$^{-1}$,
$r_{ref}$ and $r_e$ in $\AA$, $a_i$ in ${\AA}^{-1}$, $p$ is the
dimensionless. $T_{dis}$, $r_{ref}$ and $p$ parameters were fixed in
the fit.} \label{Abpar}
\begin{center}
\begin{tabular}{ccc}
\hline \hline Parameter & $A^1\Sigma^+$ & $b^3\Pi_0$\\
\hline $T_{dis}$ & \multicolumn{2}{c}{15616.95}\\
\hline
$p$ &  3  &  4\\
$r_{ref}$ &  5.0  &  4.2\\
\hline
${\mathfrak D}_e$ & 5567.546 & 6599.300\\
$r_e$   &  4.981380 &  4.179865\\
\hline
$a_0$               &  0.44685377 &  0.56383223\\
$a_1$               &  0.01153475 &  0.11731134\\
$a_2$               &  0.01224621 &  0.10163399\\
$a_3$               &  0.12946458 & -0.04095288\\
$a_4$               &  0.16407832 & -0.37280647\\
$a_5$               &  0.28366171 &  0.11826773\\
$a_6$               & -0.27382905 &  2.22890307\\
$a_7$               & -0.61363449 &  0.57576966\\
$a_8$               &  1.05182596 & -5.39299688\\
$a_9$               &  0.87045466 & -3.16111671\\
$a_{10}$            & -1.00928649 &  8.86174318\\
$a_{11}$            & -0.72414478 &  4.22763021\\
$a_{12}$            &  0.00055024 & -7.36542364\\
$a_{13}$            &             & -0.00010791\\
$a_{14}$            &             &  0.12543766\\
\hline
\end{tabular}
\end{center}
\end{table}

%******************************** Table II ********************************
\begin{table}[t!]%[H]
\caption{The resulting EMO parameters of both diagonal
$A^{so}_{\pm}$ and off-diagonal $\xi^{so}_{Ab0}$ spin-orbit coupling
functions.  $^{\dag}\xi_{Cs}^{so}$, ${\mathfrak D}_e^{so}$ and
$T_e^{so}$ in cm$^{-1}$, $^{\dag}r_{ref}$ and $r_e^{so}$ in $\AA$,
$a_i$ in ${\AA}^{-1}$, $^{\dag}p=1$ is the dimensionless. $^{\dag}$
denotes the fixed parameter. The empirical $r$-independent coupling
parameter $\zeta_{Ab1}=0.04935$.}\label{SOpar}
\begin{center}
\begin{tabular}{cccc}
\hline \hline
Parameter & $\xi^{so}_{Ab0}$ & $A^{so}_{-}$ & $A^{so}_{+}$\\
\hline
\hline $^{\dag}\xi_{Cs}^{so}$ & \multicolumn{3}{c}{184.68}\\
%$^{\dag}p$ & \multicolumn{3}{c}{1}\\
$^{\dag}r_{ref}$ &  5.3 & \multicolumn{2}{c}{5.5}\\
\hline
${\mathfrak D}_e^{so}$ & 102.998   & 106.419 & 101.539\\
$r_e^{so}$ & 5.054338  & 5.442732  & 5.576276\\
$a_0$ & 0.31895   & 0.42157   & 0.37735\\
\hline
$^{\dag}a_1$ & 0.40997 & \multicolumn{2}{c}{1.02228}\\
$^{\dag}a_2$ & 0.49244 & \multicolumn{2}{c}{2.23836}\\
$^{\dag}a_3$ &         & \multicolumn{2}{c}{-1.71726}\\
$^{\dag}a_4$ &         & \multicolumn{2}{c}{-9.84181}\\
\hline $T_e^{so}$ & 81.68  & 78.26 & 83.14\\
\hline\hline
\end{tabular}
\end{center}
\end{table}

%******************************** Table III ********************************
\begin{table}%[t!]%[H]
\caption{Comparison of the basic spectroscopic constants available
for the KCs $A^1\Sigma^+$ and $b^3\Pi$ states. The $T_e$ and
$\omega_e$ values are given in cm$^{-1}$ while $r_e$ in \AA. $T_e$
values are reffered to the minimum of the $X$ state. The harmonic
frequency for the present PECs $U(r)$ is calculated as $\omega_e=
\hbar\sqrt{k/\mu}$ \cite{field}, where $k=d^2U/dr^2|_{r=r_e}$ is the
strength constant.}\label{harm}
\begin{center}
\begin{tabular}{cccccccccc}
\hline \hline
State & $T_e$ & $r_e$ & $\omega_e$ & $T_e$ & $r_e$ & $\omega_e$& $T_e$ & $r_e$ & $\omega_e$\\
\hline & \multicolumn{3}{c}{present} &
\multicolumn{3}{c}{\cite{korek-06}}\\
\hline
$A^1\Sigma^+$ & 10049 & 4.98 & 49.8 \\
$b^3\Pi_0$    &  8833 & 4.18 & 68.4 & 8717& 4.17& 70.3\\
$b^3\Pi_1$    &  8938 & 4.19 & 68.3 & 8856& 4.16& 71.6\\
$b^3\Pi_2$    &  9044 & 4.20 & 68.2 & 8981& 4.15& 71.8\\
\hline& \multicolumn{3}{c}{\cite{korek-00}} &
\multicolumn{3}{c}{\cite{Aymar2008}}&
\multicolumn{3}{c}{\cite{KS-09}}\\ \hline
$A^1\Sigma^+$ & 10107& 4.93 & 51.8 & 9947& 4.95& 51.8& 10203& 5.02& 48.3\\
$b^3\Pi_1$    &  9017& 4.16 & 71.6 & 8870& 4.19& 69.9&  9049& 4.21& 68.8\\
\hline
\end{tabular}
\end{center}
\end{table}

\begin{figure}[t!]
\includegraphics[scale=1.0]{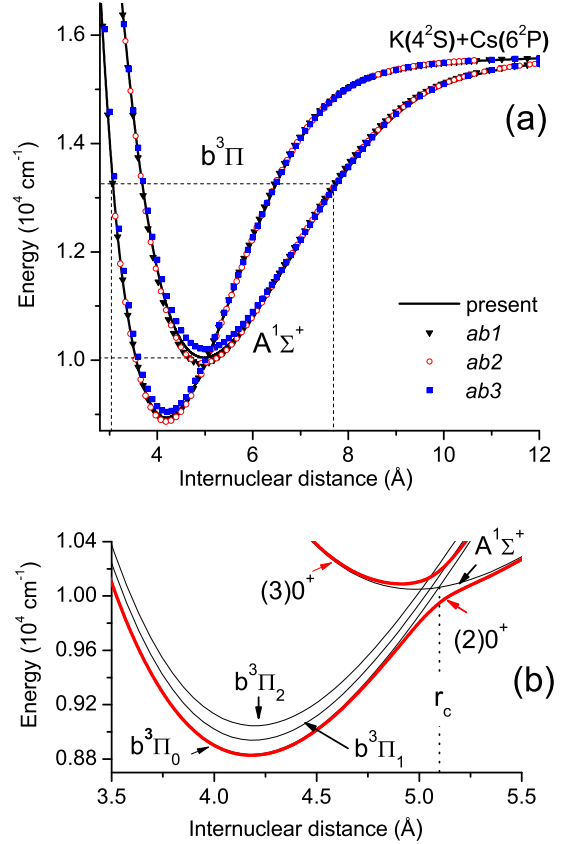}
\caption{(Color online) (a) The present empirical (\emph{emp.}) and
available \emph{ab initio} (\emph{ab1}-\cite{korek-00};
\emph{ab2}-\cite{Aymar2008}; \emph{ab3}-\cite{KS-09}) PECs of the
$A^1\Sigma^+$ and $b^3\Pi_{\Omega=1}$ states corresponding to the
\emph{a} Hund's coupling case. The insert zooms in the respective
PECs around $r_e$. (b) The present empirical diabatic PECs of the
$A^1\Sigma^+, b^3\Pi_{\Omega=0,1,2}$ sub-states and  the
corresponding adiabatic PECs $(2; 3)\Omega = 0^+$ in the \emph{c}
coupling case  states.}\label{figAbPECemp}
\end{figure}

As can be seen from Fig.\ \ref{figAbPECemp}, the singlet $A$ and
triplet $b$ state PECs are intersecting in the vicinity of the
equilibrium distance $r_e$ of the $A$-state. As a result, all levels
of the singlet state are strongly perturbed and the most significant
perturbations take place for low vibrational levels $v_A$ of the
$A$-state. The diabatic (deperturbed) $v_{b0}=18$ vibrational level
of the triplet state is found to be the first one which reach by
energy the ground diabatic $v_A=0$ level of the singlet at low
$J^{\prime}$-values. However high rotational levels of the $v_A=0$
basically interact with lower vibrational levels ($v_{b0}< 18$) of
the $b$-triplet since $r_e^b \ll r_e^A$ (see Table \ref{harm}).

\begin{figure}[t!]
\includegraphics[scale=0.7]{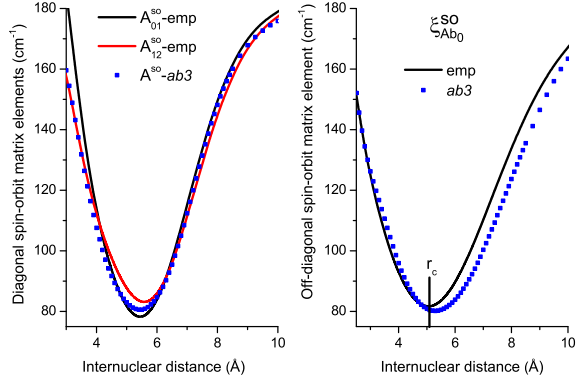}
\caption{(Color online) Comparison of the present empirical first
order spin-orbit functions with their \emph{ab initio} counterparts
(\emph{ab}) from Ref.\ \cite{KS-09}. (a) the diagonal empirical
$A^{so}_{\pm}(r)$ and \emph{ab initio} $A^{so}(r)$ SO splitting
functions; (b) the relevant off-diagonal $\xi^{so}_{Ab0}(r)$ SO
coupling functions. $r_c$ is the internuclear distance where
diabatic PECs intersect.}\label{figso}
\end{figure}

Fig.\ \ref{figAbPECemp} and \ref{figso} demonstrate a good agreement
of the derived PECs and SO matrix elements with the \emph{ab initio}
quasi-relativistic results \cite{korek-00,Aymar2008,KS-09} both
inside and outside the experimental data region $r\in [3.1,7.8]$
\AA. In particular, Table \ref{harm} clearly shows that the
empirical molecular constants $T_e$, $\omega_e$ and $r_e$ coincide
with their \emph{ab initio} counterparts within 1-2\%. The
arithmetic mean of the empirical diagonal SO matrix elements
$(A^{so}_{-}+A^{so}_{+})/2$ (see Fig.\ \ref{figso}a) agrees with the
\emph{ab initio} SO splitting $A^{so}$ from Ref.\ \cite{KS-09}
within few cm$^{-1}$ near a minimum of these functions; (see Eq.\
(\ref{Aso1}). However, the present empirical $A^{so}_{\pm}(r)$
functions evaluated near the equilibrium distance $r_e$ of the
triplet $b$-state are smaller on ~20-25 cm$^{-1}$ than their
\emph{ab initio} counterparts corresponding to the pure (\textbf{c})
Hund's coupling case \cite{korek-06} (see Table \ref{harm}). Near
the crossing point of singlet and triplet states $r_c\approx 5.1$
\AA (Fig.\ \ref{figAbPECemp}) the empirical off-diagonal SO coupling
function $\xi^{so}_{Ab0}$ coincide with \emph{ab initio} result
\cite{KS-09} within $1-2$ cm$^{-1}$ (Fig.\ \ref{figso}b).

\begin{figure}[t!]
\includegraphics[scale=1.0]{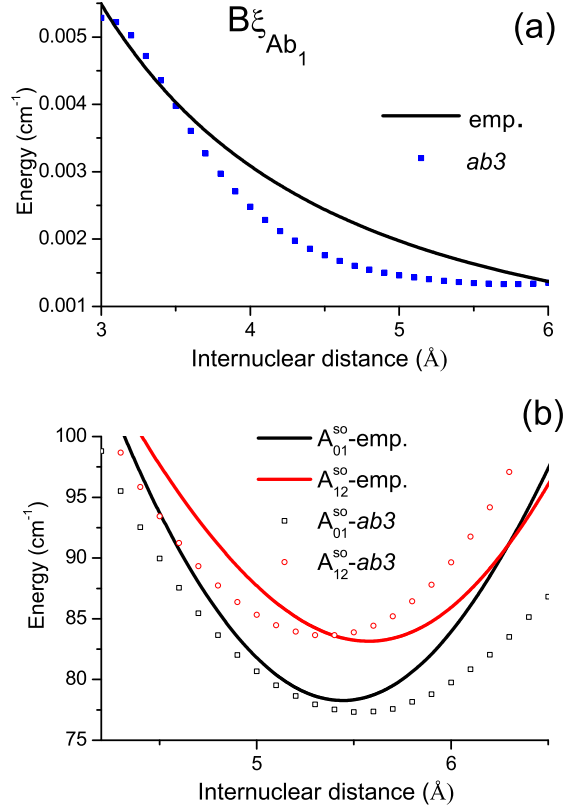}
\caption{Comparison of the present empirical and \emph{ab initio}
second order spin-orbit functions. (a) the empirical $B\zeta_{Ab1}$
entering Eq.\ (\ref{modelodAb4}) while \emph{ab initio} function is
determined by Eq.\ (\ref{Ab1}); (b) the empirical SO splitting
functions $A^{so}_{\pm}$ entering Eq.\ (\ref{ASO}) while \emph{ab
initio} $A^{so}_{\pm}$ values are calculated by Eq.\
(\ref{Aso1}).}\label{figso1}
\end{figure}

The empirical matrix element $B\zeta_{Ab1}$ (see Fig.\
\ref{figso1}a) reasonably agree with the \emph{ab initio} estimate
obtained in the framework of the second order perturbation theory
\begin{eqnarray} \label{Ab1}
B\zeta_{Ab1}\approx \frac{BL^{\pm}_{A-B}
\xi^{so}_{B-b}}{(U_A+U_b)/2-U_B}
\end{eqnarray}
by using of the \emph{ab initio} PEC for the $B^1\Pi$ state $U_B$ as
well as SO $\xi^{so}_{B-b}$ and angular $L^{\pm}_{A-B}$ coupling
matrix elements \cite{KS-09}. Furthermore, the empirical
non-equidistant SO splitting of the $b^3\Pi$-state (Fig.\
\ref{figso} and \ref{figso1}) qualitatively agree with the estimates
\begin{eqnarray} \label{Aso1}
A^{so}_{\pm}=A^{so}\pm\delta A^{so};\quad \delta A^{so}\approx
\sum_{j} \frac{|\xi^{so}_{b-j}|^2}{U_b-U_j}
\end{eqnarray}
where the required PECs $U_j$ for the $j\in a^3\Sigma^+;
c^3\Sigma^+; B^1\Pi$ states and relevant SO matrix elements
$A^{so}$, $\xi^{so}_{b-j}$ of the $b^3\Pi$ state were taken from
Ref.\ \cite{KS-09}.

\begin{figure}[t!]
\includegraphics[scale=0.9]{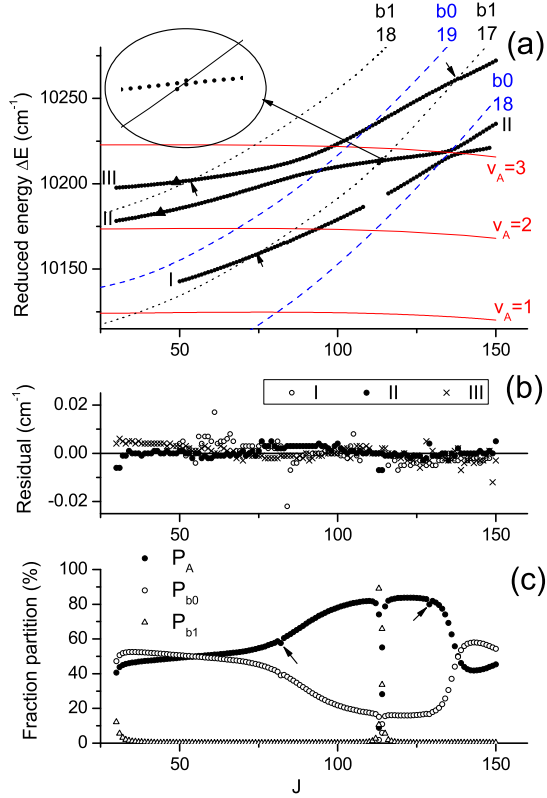}
\caption{ (a) Fragment of $J^{\prime}$-dependence of the
experimental term values as well as their deperturbed diabatic
singlet and triplet counterparts. Termvalues are given by the
reduced scale: $\Delta E=E-0.0222\times J^{\prime}(J^{\prime}+1)$.
The graphs I, II, and III denote experimental data. Almost
horizontal solid lines refer to diabatic singlet $A^1\Sigma^+$ state
in the region $v_A = 1 - 3$; the dashed and dotted lines refer to
diabatic triplet sub-states $b^3\Pi_{\Omega=0}$ ($v_{b0}=18-20$) and
$b^3\Pi_{\Omega=1}$ ($v_{b1} = 17$ and $18$), respectively. For
small $J^{\prime}$ the effective vibrational quantum numbers can be
attributed as $v^*_{b0}$(I) = 20, $v^*_{b0}$(II) = 21 and
$v^*_{A}$(III) = 1. The inset zooms in the splitting caused by the
interaction with $v_{b1} = 17$ term. Arrows mark places of observed
weak local perturbations with $b1$ sub-state. The triangles mark the
levels which give the LIF progressions analyzed in Fig.\
\ref{Intensity_v2singlet} and \ref{Intensity_J158}. (b) Residuals of
the fit $E^{expt}_j-E^{CC}_j$ for all respective $J^{\prime}$
values. (c) Fraction partition of the relevant non-adiabatic
wavefunctions $P_i$ for the experimental term values presented by
graph II. The sharp resonance at $J^{\prime}=113/114$ corresponds to
the intersection shown by inset in (a). The arrows in (c) mark the
predicted influence of $b^3\Pi_{\Omega=2}$ component.
}\label{figTermJ113}
\end{figure}

\begin{figure}[t!]
\includegraphics[scale=0.9]{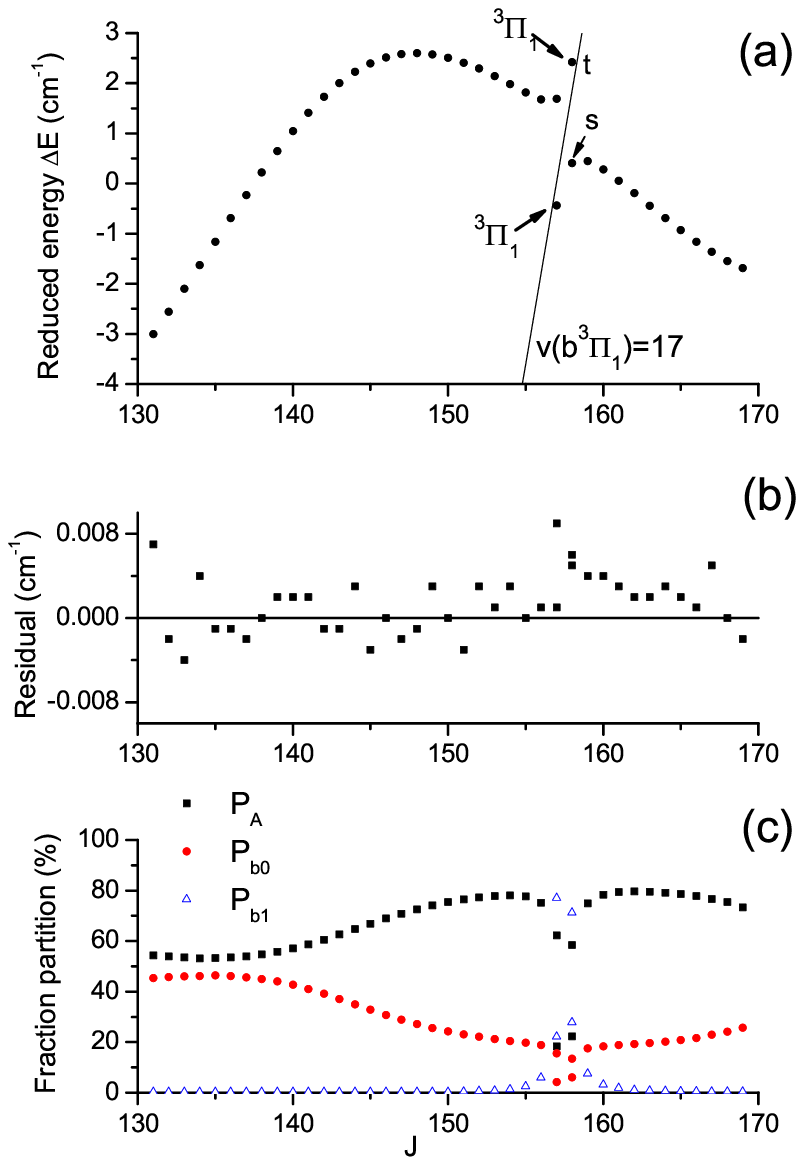}
\caption{(Color online) (a) The experimental termvalues around the
local $J^{\prime}=157/158$ perturbation (see Fig.\ \ref{loc_pert}b)
caused by the $v_{b1}=17$ level of the diabatic $b^3\Pi_{\Omega=1}$
sub-state (marked by the solid line). Termvalues are presented by
the reduced scale: $\Delta E=E-10266-0.0237
J^{\prime}(J^{\prime}+1)$. (b) Residuals of the fit
$E^{expt}-E^{CC}$ for respective $J^{\prime}$ values. (c) The
respective partition functions $P_i$.}\label{figTermJ158}
\end{figure}

\subsection{Details of the $A^1\Sigma^+$-state interaction with
the $b^3\Pi_{\Omega=0,1}$ sub-states}

The magnitude of the off-diagonal \emph{homogenous} spin-orbit
$A^1\Sigma^+\sim b^3\Pi_{\Omega=0}$ interaction, see Fig.\
\ref{figso}, is comparable with vibrational spacing of the
interacting states, see Table \ref{harm}. This leads to significant
regular non-adiabatic shifts of the most experimental level
positions of the $A\sim b$ complex with respect to their deperturbed
(diabatic) singlet and triplet counterparts. Fig.\
\ref{figTermJ113}a gives us a typical example of the experimental
level positions of the complex given in a reduced scale as function
of rotational quantum number $J^{\prime}$. As can be seen, the
experimental $J^{\prime}$-dependence of $A$ state vibrational level
identified for small $J^{\prime}$ as $v_A^*=1$ (depicted by graph
III), is shifted towards higher energies by approximately 75
cm$^{-1}$, or by about 1.5$\omega_e^A$ (compare with diabatic $v_A =
1$). The graphs I - II belong to experimental $v_{b0}^*=20$ and
$21$, respectively. The experimental levels are significantly
shifted down with respect their dibatic (deperturbed) positions. All
three graphs I - III demonstrate a different slope from any of
diabatic $J^{\prime}$-dependencies. Moreover, the slope is changing
with $J^{\prime}$ demonstrating the singlet-triplet anti-crossing
effect which takes place between experimental terms II and III
around $J^{\prime}=80$ while between I and II around $J^{\prime}=
130$.

Along with the dominating $A^1\Sigma^+\sim b^3\Pi_{\Omega=0}$
interaction, weak local \emph{heterogenous} perturbations caused by
the $(A-b)_{\Omega=0}\sim b^3\Pi_{\Omega=1}$ indirect interaction
are pronounced as sharp irregularities of the order of 1 cm$^{-1}$
in the vicinity of intersection of experimental term values with
diabatic dependencies of the $b^3\Pi_{\Omega=1}$ sub-state. One such
example is given by the inset in Fig.\ \ref{figTermJ113}a while the
more detailed analysis of the \emph{heterogenous} perturbation is
shown in Fig.\ \ref{figTermJ158}. The peculiarity of the both cases
is connected with experimental observation of two pairs of lines
with the same $J^{\prime}$-value, namely, for $J^{\prime}=113/114$
and $J^{\prime}= 157/158$ (see Fig.\ \ref{loc_pert}b for the second
example), giving levels with slightly different energies. This is
explained by a local perturbation, as a result of which, in each
pair of levels with the same $J^{\prime}$, see Fig.\
\ref{figTermJ158}a and the inset in Fig.\ \ref{figTermJ113}a, one
level is predominantly singlet $A$ state, while another one is the
triplet $b^3\Pi_{ \Omega=1}$ sub-state. It should be noted that at
intersection of graphs I and III with diabatic $b_1$ graph in Fig.\
\ref{figTermJ113}a only small shifts were observed, without
appearance of pairs with the same $J^{\prime}$. This typical
situation should be attributed to too small fraction of singlet
wavefunction in the partner level to observe it.

Both types of interaction are described remarkably well by the
present deperturbation model, as is demonstrated by the residuals
between the fitted and measured termvalues, see, for example, Fig.\
\ref{figTermJ113}b and Fig.\ \ref{figTermJ158}b. Furthermore, the
calculated fraction partitions $P_i$ (see Fig.\ \ref{figTermJ113}c,
Fig.\ \ref{figTermJ158}c, Table \ref{isotope} and EPAPS
\cite{EPAPS}) unambiguously prove a dominant role of the direct SO
$A^1\Sigma^+\sim b^3\Pi_{\Omega=0}$ interaction. Indeed, one can see
that smooth variation of the singlet fraction in the total wave
function corresponds to the smooth change of the respective
termvalues. The sharp resonances at $J^{\prime}=113/114$ (Fig.\
\ref{figTermJ113}c) and $J^{\prime}=157/158$ (Fig.\
\ref{figTermJ158}c) both correspond to the intersection of
$J^{\prime}$-dependence of the respective experimental tervalues
with diabatic dependence for $v_{b1}=17$ caused by indirect
interaction of $A$ state with the $b^3\Pi_{\Omega=1}$ sub-state. The
very small irregularities of partition fractions noticed on Fig.\
\ref{figTermJ113}c at $J^{\prime}=81$ and 124 indicate the predicted
influence of $b^3\Pi_2$ component which has only minor effect on
energy and, hence, it could not be observed experimentally.

\subsection{Term values of the $^{41}$K$^{133}$Cs isotopomer}
To confirm mass-invariant properties of the deperturbed parameters
the rovibronic termvalues of the $^{41}$K$^{133}$Cs isotopomer have
been predicted and compared with the experimentally observed ones.
In the calculation only corresponding reduced mass $\mu$ of the
isotopomer in the operator kinetic energy Eq.\ (\ref{CC}) and
parameter $B$ in the Eq.\ (\ref{BJ}) of the potential energy matrix
defined by Eq.\ (\ref{modeldAb4}) and (\ref{modelodAb4}) was
substituted. The calculated termvalues coincide with the
experimental data (see Table \ref{isotope}) with standard deviation
of 0.0055 cm$^{-1}$. In absence of any adjustment of the parameters,
the agreement even for the highly excited and strongly mixing levels
can be assumed as excellent.

\subsection{Intensity distribution in the $A^1\Sigma^+\sim b^3\Pi
\to X^1\Sigma^+$ LIF progressions}\label{intensity}

As well-known a nodal structure of non-adiabatic vibrational
wavefunctions is rather sensitive to strong intramolecular
perturbations \cite{field,Docenko-2007,Zaharova-2009}. Therefore, a
comparison of the experimental relative intensity distributions in a
band structure $A\sim b\to X(v_X)$ LIF progressions with their
theoretical counterparts would be a critical independent test on a
completeness of the above energy-based deperturbation analysis.

The theoretical transition probabilities $I_{A\sim b\to X}$ were
estimated as:
\begin{eqnarray}\label{Iten}
I_{A\sim b\to X}(v_X)\sim \nu^4_{A\sim b\to X}M^2_{A\sim b-X}\\
\nu_{A\sim b -X}=E^{CC}(J^{\prime})-E_X(v_X;J_X=J^{\prime}\pm 1)\nonumber\\
M_{A\sim b-X} =\langle\phi_A|d_{AX}|v_X\rangle\nonumber
\end{eqnarray}
where $d_{AX}(r)$ is the \emph{ab initio} $A - X$ transition dipole
moment from Ref's \cite{Aymar2008, KS-09}. The power of $4$ used in
the transition wavenumber $\nu_{A\sim b\to X}$ assumes that the
detector is proportional to the intensity of the incoming
fluorescence light. The rovibronic energies $E_X$ and eigenfunctions
$|v_X\rangle$ of the ground $X^1\Sigma^+$ state were obtained by
solving the single channel radial equation with the empirical PEC
from Ref.\ \cite{KCs-ground}. Hereafter, both experimental and
calculated intensities obtained for $P$ and $R$-branches separately
were then averaged since they basically were very close to each
other.

\begin{figure}[t!]
\includegraphics[scale=1.0]{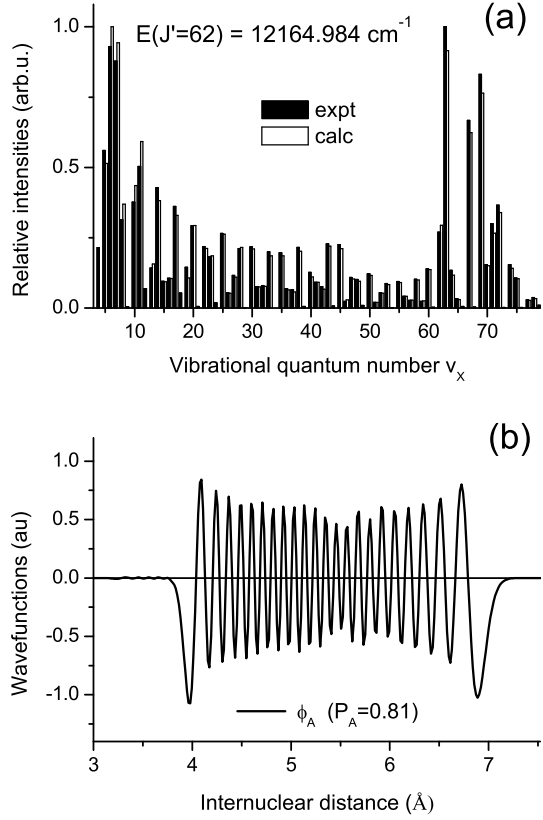}
\caption{(a) The experimental and calculated intensity distribution
in the long LIF progression (see Fig.\ \ref{spec1}) originating from
the high rovibronic level $E_{A \sim b}(J^{\prime}=62)$ of the $A
\sim b$ complex. (b) Respective fraction of the non-adiabatic
$\phi_A(r)$ wavefunction of the $A$-state ($P_A=\langle\phi_A|\phi_A
\rangle=0.81$).} \label{Intensity_long}
\end{figure}

The remarkable agreement observed (see Fig.\ \ref{Intensity_long}a)
between the simulated and experimental intensity distribution in the
very long $v_X\in [0,80]$ LIF progression originating from the high
rovibronic level of the complex convincingly proves a high
reliability of the deperturbation model used. It is easily seen from
Fig.\ \ref{Intensity_long}b that the non-adiabatic wavefunction
$\phi_A(r)$ of the singlet $A$-state, which is responsible for the
complicated oscillation behavior of intensity observed on Fig.\
\ref{Intensity_long}a, is only weekly perturbed by the triplet
$b$-state. The respective fraction of the singlet state $P_A$ is
dominating here being 0.81.

\begin{figure}[t!]
\includegraphics[scale=0.9]{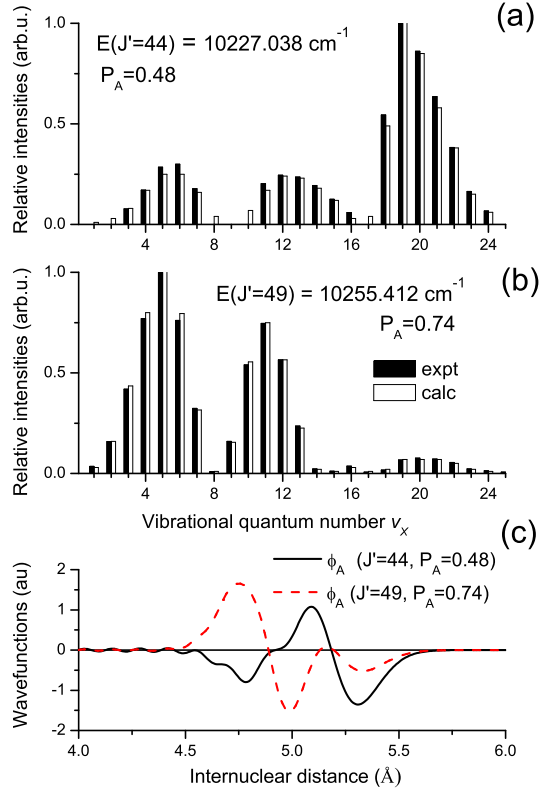}
\caption{The experimental and calculated intensity distributions in
LIF progressions originating from close-lying rovibronic levels of
the $A\sim b$ complex (marked by the solid triangles in Fig.\
\ref{figTermJ113}): (a) for the $v^*_{b0}=21; J^{\prime}=44$ having
an admixture of the triplet $b^3\Pi_{\Omega=0}$ component
($P_{b0}=\langle\phi_{b0}| \phi_{b0}\rangle=0.52$); (b) for the
$v^*_A=1; J^{\prime}=49$ level having a dominant singlet
$A^1\Sigma^+$ character ($P_A=0.74$). (c) Respective non-adiabatic
wavefunctions $\phi_A(r)$ of the singlet
$A$-state.}\label{Intensity_v2singlet}
\end{figure}

To elucidate influence of the strong off diagonal homogenous SO
$A^1\Sigma^+\sim b^3\Pi_{\Omega = 0}$ interactions on a nodal
structure of the perturbed wavefunctions we have measured and
simulated respective intensity distributions in $A\sim b\to X(v_X)$
LIF progressions originating from two rovibronic levels (marked by
solid triangles on Fig.\ \ref{figTermJ113}a) which are close to each
other by $J^{\prime}$ and term values. These levels have, however, a
quite different admixture of the $b^3\Pi_0$ component. The agreement
between calculated and experimental intensities is found to be, once
more, remarkably good (see Fig.\ \ref{Intensity_v2singlet}a,b). For
both levels the $I(v_X)$ functions (Fig.\
\ref{Intensity_v2singlet}a,b) obey very well to so-called ``Condon
\emph{reflection} approximation" \cite{field} since the $v_X$
dependence of $I(v_X)$ mimic the $r$-dependence of the initial
wavefunction of the complex, $|\phi_A(r)|^2$; see Fig.\
\ref{Intensity_v2singlet}c. Indeed, in according to three lobes of
non-adiabatic wavefunctions $\phi_A(r)$ of singlet fraction depicted
on Fig.\ \ref{Intensity_v2singlet}c the respective intensity
distributions show three pronounced maximums which are clearly seen
on Fig.\ \ref{Intensity_v2singlet}a,b. At the same time, due to the
strong homogenous interaction an amplitude of the lobes of two
wavefunctions are well distinguished leading to quiet different
intensity distribution functions (compare Fig.\
\ref{Intensity_v2singlet}a with Fig.\ \ref{Intensity_v2singlet}b).
It is interesting that the most dramatic changes in a shape of
non-adiabatic wavefunctions are observed near the crossing point
$r_c$ of the interacting states (Fig.\ \ref{figAbPECemp}b). For
instance, the $\phi_A(r)$ of the $J^{\prime}=44$ level has very
small additional lobe near $r\approx 5.2$ \AA. This lobe provides
small but not negligible peak located at $v_X=16$ on the intensity
distribution function; see Fig.\ \ref{Intensity_v2singlet}b.

It also should be noticed that in according to ``\emph{oscillation}
theorem" \cite{landau} the $J^{\prime}=49$ level mentioned above
would be assigned to the $v_A=2$ vibrational state since its
wavefunction $\phi_A(r)$ has two nodes at least. However, from the
energy viewpoints (Fig.\ \ref{figTermJ113}a) and $P_A$ analysis it
can be assigned as $v_A^*=1$. The detailed analysis of this effect
(conventional \emph{oscillation} theorem breakdown caused by strong
homogeneous perturbations) has been done separately in our
accompanied paper \cite{OT}.

\begin{figure}[t!]
\includegraphics[scale=0.9]{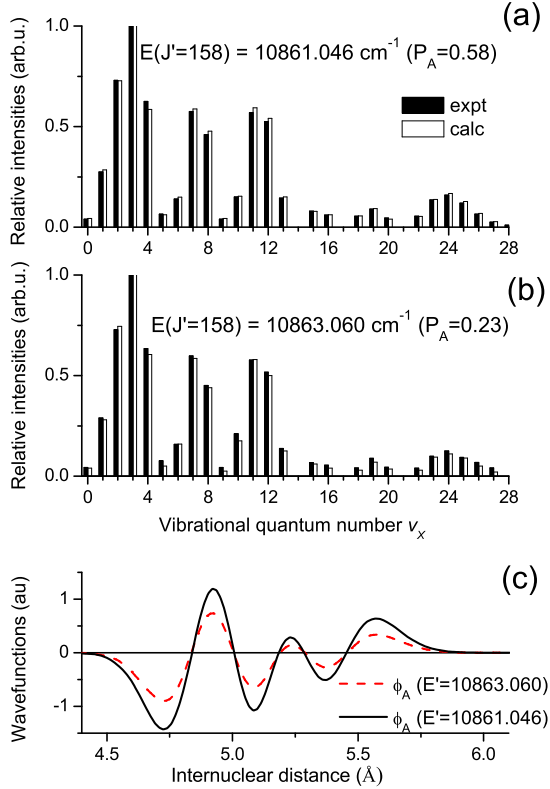}
\caption{The experimental and calculated intensity distributions in
the LIF progressions originating from two mutually perturbed levels
of the $A\sim b$ complex with the same $J^{\prime}=158$-value (see
Fig.\ \ref{figTermJ158}a): (a) having dominant singlet $A^1\Sigma^+$
character ( $P_A^s=0.58$; $P_{b0}^s=0.14$; $P_{b1}^s=0.28$ ), marked
in Fig.\ \ref{figTermJ158}a by ``\textbf{s}"; (b) having dominant
$b^3\Pi_{\Omega=1}$ component ($P_A^t=0.23$; $P_{b0}^t=0.06$;
$P_{b1}^t=0.71$ ), marked by ``\textbf{t}" in Fig.\
\ref{figTermJ158}a. (c) Respective non-adiabatic $\phi_A^{s/t}(r)$
wavefunctions of the $A$-state.}\label{Intensity_J158}
\end{figure}

To investigate the influence of the local \emph{heterogenous}
$(A-b)_{\Omega=0}\sim b^3\Pi_{\Omega=1}$ interaction on a nodal
structure of wavefunctions we have measured and simulated intensity
distributions in the LIF progressions originating from two mutually
perturbed levels, shown on Fig.\ \ref{figTermJ158}a, with the same
$J^{\prime}=158$ values. The agreement between calculated and
experimental values, once more, is remarkably good (Fig.\
\ref{Intensity_J158}a,b). It is surprising to see that the intensity
distributions for both levels are found to be almost identical; the
one level (marked as ``s") having the lower energy demonstrates a
dominant singlet $A^1\Sigma^+$ character while the higher-lying
level (marked as ``t") has significant $b^3\Pi_{\Omega=1}$
component. The calculation of the respective non-adiabatic
wavefunctions $\phi_A^{s/t}(r)$ (see Fig.\ \ref{Intensity_J158}c)
confirms that their nodal structure is coinciding while the
amplitudes are different. Indeed, the wavefunctions relates to each
other as
\begin{eqnarray}
\sqrt{P_A^t}\phi_A^{s}(r)=\sqrt{P_A^s}\phi_A^{t}(r) .
\end{eqnarray}

\subsection{Simulation of the optical cycle $a^3\Sigma^+ \to A^1\Sigma^+\sim b^3\Pi \to
X^1\Sigma^+$}\label{raman}

Since the predictive abilities of the derived deperturbation
parameters of the $A\sim b$ complex were unambiguously proved above
we have used them to simulate a wavenumbers
\begin{eqnarray}\label{wavenumber}
\nu^{PUMP}_{a\to A\sim b}=E^{CC}_j(J^{\prime}=1)-E_a(N_a=0)\nonumber\\
\nu^{DUMP}_{A\sim b \to
X}=E^{CC}_j(J^{\prime}=1)-E_X(v_X=0;J_X=0)\nonumber
\end{eqnarray}
and transition moments
\begin{eqnarray}\label{tran}
M^{PUMP}_{A\sim b-a}=|\langle\phi_{b_{\Omega}}|d_{ba}|\chi_a\rangle|\nonumber\\
M^{DUMP}_{A\sim b-X}=|\langle\phi_A|d_{AX}|\chi_X\rangle|\nonumber
\end{eqnarray}
for the PUMP-DUMP optical cycle $a^3\Sigma^+\to A^1\Sigma^+\sim
b^3\Pi \to X^1\Sigma^+$ which has been proposed in Ref.\
\cite{stw-04} to transform an ultracold K(3$^2$S)$+$Cs(6$^2$S)
colliding pairs in their absolute ground state. The excitation $a\to
A\sim b$ and emission $A\sim b \to X$ transitions presented in Table
\ref{SRPD} correspond to the most pronounced transition
probabilities of the two-steps conversion process. Table \ref{SRPD}
also contains the predicted radiative lifetimes $\tau_{A\sim b}$
\cite{JCP03}
%\begin{widetext}
\begin{eqnarray}\label{tausum}
\frac{1}{\tau_{A\sim b}} = \frac{8\pi^2}{3\hbar c}\times \\
\left[\langle\phi_A|\Delta U_{AX}^3d_{AX}^2|\phi_A\rangle +
\langle\phi_b|\Delta U_{ba}^3d_{ba}^2|\phi_b\rangle\right]\nonumber
\end{eqnarray}
%\end{widetext}
and branching ratios of the spontaneous emission $R_{A\sim b}$
\begin{eqnarray}\label{branchratio}
R_{A\sim b} =\frac{8\pi^2}{3\hbar c} \nu^3_{A\sim b -X}M^2_{A\sim
b-X}\tau_{A\sim b}
\end{eqnarray}
from the intermediate level of the complex with $J^{\prime}=1$ to
the absolute ground level. Here, $\Delta U_{ij}=U_i-U_j$ is the
difference potential. Both $\tau_{A\sim b}$ and $R_{A\sim b}$ values
are also useful for estimating the efficiency of the simple
PUMP-spontaneous emission process for producing population in the
absolute ground level.

The probabilities of PUMP transition were estimated using the
\emph{ab initio} dipole transition moment $d_{ba}(r)$ from Ref.\
\cite{KS-09} and adiabatic wavefunction $|v_a\rangle$ of the initial
$a^3\Sigma^+$ state \cite{KCs-triplet} calculated for a virtual
vibrational level with the zero bounding energy $E_a(N=0)=0$. The
singlet-triplet $X^1\Sigma^+ \sim a^3\Sigma^+$ perturbation of the
initial level caused by a hyperfine Fermi contact interaction
\cite{KCs-triplet} was neglected. The upper limit of the termvalues
for the $A\sim b$ complex $E^{CC}_j(J^{\prime}=1)$ was limited by
13200 cm$^{-1}$ in order to avoid an extrapolation error outside the
experimental region. The wavenumbers and transition probabilities
for arbitrary rovibronic levels of the $A\sim b$ complex and ground
singlet $X$ and triplet $a$ states can be generated by a request.

%******************************** Table III ********************************
\begin{table}[t!]
\caption{The wavenumbers $\nu^{PUMP}_{a\to A\sim b}$,
$\nu^{DUMP}_{A\sim b \to X}$ (in cm$^{-1}$) and transition moments
$M^{DUMP}_{A\sim b-X}$, $M^{DUMP}_{A\sim b-a}$ (in $a.u.$) predicted
for the most favorable stimulated Raman process $a^3\Sigma^+(N_a=0)
\to A^1\Sigma^+\sim b^3\Pi(J^{\prime}=1) \to
X^1\Sigma^+(v_X=0;J_X=0)$. The $\nu^{PUMP}_{a\to A\sim b}$ values
are given with respect to the common dissociation energy of the
ground singlet $X$ and triplet $a$ states (see Fig.1). The
$\tau_{A\sim b}$ (in $ns$) and $R_{A\sim b}$ (in $\%$) are the
radiative lifetimes and branching ratios of the spontaneous
emission, respectively.}\label{SRPD}
\begin{center}
\begin{tabular}{cccccc}
\hline \hline $\nu^{PUMP}_{a\to A\sim b}$ & $\nu^{DUMP}_{A\sim b \to
X}$ & $M^{PUMP}_{A\sim b-a}$ & $M^{DUMP}_{A\sim b-X}$  &
$\tau_{A\sim b}$ & $R_{A\sim b}$ \\
\hline
7375.17 & 11410.643 & 1.03(-2) & 0.189&  61 & 1\\
7311.92 & 11347.386 & 5.89(-3) & 0.223&  83 & 1\\
7182.31 & 11217.785 & 4.06(-3) & 0.499&  40 & 3\\
7121.44 & 11156.906 & 7.48(-3) & 0.453&  69 & 4\\
7100.85 & 11136.325 & 5.74(-3) & 0.528&  40 & 3\\
7047.54 & 11083.010 & 7.77(-3) & 0.795&  35 & 6\\
7018.76 & 11054.231 & 6.02(-3) & 0.326&  73 & 2\\
6990.99 & 11026.460 & 5.08(-3) & 0.674&  52 & 6\\
6967.69 & 11003.164 & 6.66(-3) & 0.818&  44 & 8\\
6910.53 & 10946.001 & 3.66(-3) & 1.117&  34 & 11\\
6834.33 & 10869.801 & 1.54(-3) & 1.086&  53 & 16\\
6811.20 & 10846.672 & 2.56(-3) & 0.486&  62 & 4\\
6771.69 & 10807.165 & 3.45(-3) & 1.352&  34 & 16\\
6764.03 & 10799.497 & 6.12(-3) & 0.375& 117 & 4\\
6700.45 & 10735.916 & 9.81(-3) & 1.224&  59 & 22\\
6630.94 & 10666.413 & 7.46(-3) & 1.408&  35 & 17\\
6590.33 & 10625.805 & 3.72(-3) & 0.512&  59 & 4\\
6564.47 & 10599.938 & 7.16(-3) & 1.092&  54 & 15\\
6544.13 & 10579.603 & 3.40(-3) & 0.789&  46 & 7\\
6488.40 & 10523.869 & 2.45(-3) & 1.104&  34 & 10\\
6343.98 & 10379.452 & 5.38(-3) & 0.613&  34 & 3\\
6268.13 & 10303.598 & 8.10(-3) & 0.338&  43 & 1\\
\hline
\end{tabular}
\end{center}
\end{table}

\section{Conclusions}
We have accomplished both experimental and deperturbation studies of
the fully mixed $A^1\Sigma^+$ and $b^3\Pi$ states of the KCs
molecule based on a direct reduction of highly accurate
Fourier-transform spectroscopy data on rovibronic termvalues of the
$A\sim b$ complex to potential energy curves and spin-orbit coupling
functions for the mutually perturbed states.

The experimental data field starts from the lowest vibrational level
$v^*_A=0$ of the singlet and nonuniformly covers rotational quantum
numbers $J\in [7,225]$ in the energy range $E^J\in [10040,13250]$
cm$^{-1}$. Overall 42 adjusted fitting parameters have been required
to reproduce more than 3400 rovibronic termvalues of the $A\sim b$
complex with a standard deviation of 0.004 cm$^{-1}$ which is
absolutely consistent with the uncertainty of the experiment of
0.003-0.01 cm$^{-1}$. Reliability of the derived structure
parameters was unambiguously confirmed by a good agreement of the
predicted termvalues of $^{41}$K$^{133}$Cs isotopomer and relative
intensity distributions in the $A\sim b \to X(v_X)$ LIF progressions
with their experimental counterparts.

Besides of the dominating \emph{homogeneous} spin-orbit
$A^1\Sigma^+\sim b^3\Pi_{\Omega = 0}$ interactions the local
\emph{heterogenous} $A^1\Sigma^+\sim b^3\Pi_{\Omega = 1}$
perturbations have been discovered in the LIF spectra and their
impact on a nodal structure of the non-adiabatic vibrational
wavefunctions of the complex have been analyzed by means of
specially measured intensity distributions. It was found that the
\emph{homogeneous} perturbations can change dramatically a shape of
wavefunctions (including even total numbers and position of their
nodes) whereas the \emph{heterogenous} perturbations affect only on
amplitude of the wavefunction in according to its fraction
partition.

The systematic deviations greater then 0.01 cm$^{-1}$ of the
experimental and reproduced termvalues are still observed mostly for
lying higher then 12000 cm$^{-1}$ rovibronic levels of the complex
especially when they poses the significant admixture of the
$b^3\Pi_{\Omega=1}$ component. This drawback of the deperturbation
model could be attributed to the neglected influence of first order
spin-orbit and electronic-rotational interactions of the $A\sim b$
complex with nearest $c^3\Sigma^+$ state (Fig.\ \ref{PECab}).

The strong $A^1\Sigma^+ \sim b^3\Pi$ spin-orbit interaction opens a
window for direct observation of the formally spin-forbidden and
basically very weak $b^3\Pi-X^1\Sigma^+$ transitions originating
from the rovibronic levels of the $b$-state lying below $v_A = 0$ of
the perturbing $A^{1}\Sigma^+$ state. This work is in a progress.

The Table \ref{SRPD} shows that the efficiency of the proposed
conversion cycle $a^3\Sigma^+(N=0)\to A^1\Sigma^+\sim
b^3\Pi(J'=1)\to X^1\Sigma^+(v_X=0;J_X=0)$ is limited by its
absorbtion part which requires a rather high power laser source
generating in not very favorable spectral diapason. This limitation
 caused by a very weak $b-a$ triplet transition dipole
moment $d_{ba}(r)$ in the considered $r$-range \cite{Aymar2008,
KS-09}, has been discovered in the analogue cycle of NaRb
\cite{Docenko-2007} and NaCs \cite{Zaharova-2009} molecules.
Nevertheless, relative high efficiency of the single step PUMP-
spontaneous emission transitions observed for the particular levels
of the $A\sim b$ complex is encouraging for planning such
experiments in the future.

\begin{acknowledgments}
The authors are indebted to A.\ Pashov for providing the program
package for identification and analysis of LIF progressions. We are
grateful to Dr. Andrei Jarmola for providing the laser diodes built
in home made external cavity resonators and to V. Zuters for his
help in spectra analysis. The support from the Latvian Science
Council grant Nr 09.1036 is gratefully acknowledged by Riga team.
Moscow team thanks for the support of Federal Program ``Scientists
and educators for innovative Russia 2009-2013", contract P 2280.
\end{acknowledgments}

\end{document}